\pdfoutput=1

\documentclass[12pt,a4paper]{article}

\usepackage{ifthen} 
\newboolean{pdflatex}
\setboolean{pdflatex}{true} 

\newboolean{articletitles}
\setboolean{articletitles}{true} 

\newboolean{uprightparticles}
\setboolean{uprightparticles}{false} 

\newboolean{inbibliography}
\setboolean{inbibliography}{false} 

\usepackage{microtype}
\usepackage{lineno}  
\usepackage{xspace} 
\usepackage{caption} 

\usepackage{graphicx}  
\usepackage{color}
\usepackage{colortbl}
\graphicspath{{./figs/}} 

\usepackage{amsmath} 
\usepackage{amssymb}
\usepackage{amsfonts}
\usepackage{upgreek} 

\newcommand*\patchAmsMathEnvironmentForLineno[1]{%
\expandafter\let\csname old#1\expandafter\endcsname\csname #1\endcsname
\expandafter\let\csname oldend#1\expandafter\endcsname\csname
end#1\endcsname
 \renewenvironment{#1}%
   {\linenomath\csname old#1\endcsname}%
   {\csname oldend#1\endcsname\endlinenomath}%
}
\newcommand*\patchBothAmsMathEnvironmentsForLineno[1]{%
  \patchAmsMathEnvironmentForLineno{#1}%
  \patchAmsMathEnvironmentForLineno{#1*}%
}
\AtBeginDocument{%
\patchBothAmsMathEnvironmentsForLineno{equation}%
\patchBothAmsMathEnvironmentsForLineno{align}%
\patchBothAmsMathEnvironmentsForLineno{flalign}%
\patchBothAmsMathEnvironmentsForLineno{alignat}%
\patchBothAmsMathEnvironmentsForLineno{gather}%
\patchBothAmsMathEnvironmentsForLineno{multline}%
\patchBothAmsMathEnvironmentsForLineno{eqnarray}%
}

\usepackage{hyperref}    
\usepackage[all]{hypcap} 

\usepackage{xspace} 
\usepackage{upgreek}

\def\lhcb {\mbox{LHCb}\xspace}

\def\MagUp {\mbox{\em Mag\kern -0.05em Up}\xspace}

\ifthenelse{\boolean{uprightparticles}}%
{

 \def\Pmu         {\ensuremath{\upmu}\xspace}

 \def\Ppi         {\ensuremath{\uppi}\xspace}

 \def\Ppsi        {\ensuremath{\uppsi}\xspace}

 \def\PDelta      {\ensuremath{\Delta}\xspace}                 
 \def\PXi      {\ensuremath{\Xi}\xspace}                 
 \def\PLambda      {\ensuremath{\Lambda}\xspace}                 
 \def\PSigma      {\ensuremath{\Sigma}\xspace}                 
 \def\POmega      {\ensuremath{\Omega}\xspace}                 
 \def\PUpsilon      {\ensuremath{\Upsilon}\xspace}                 
 
 \def\PB      {\ensuremath{\mathrm{B}}\xspace}                 
                  
 \def\PD      {\ensuremath{\mathrm{D}}\xspace}

 \def\PJ      {\ensuremath{\mathrm{J}}\xspace}                 
 \def\PK      {\ensuremath{\mathrm{K}}\xspace}

 \def\Pb      {\ensuremath{\mathrm{b}}\xspace}                 
 \def\Pc      {\ensuremath{\mathrm{c}}\xspace}

 \def\Pi      {\ensuremath{\mathrm{i}}\xspace}

 \def\Pp      {\ensuremath{\mathrm{p}}\xspace}

}
{

 \def\Pmu         {\ensuremath{\mu}\xspace}

 \def\Ppi         {\ensuremath{\pi}\xspace}

 \def\Ppsi        {\ensuremath{\psi}\xspace}                 
                  
 \mathchardef\PDelta="7101
 \mathchardef\PXi="7104
 \mathchardef\PLambda="7103
 \mathchardef\PSigma="7106
 \mathchardef\POmega="710A
 \mathchardef\PUpsilon="7107
                  
 \def\PB      {\ensuremath{B}\xspace}                 
                  
 \def\PD      {\ensuremath{D}\xspace}

 \def\PJ      {\ensuremath{J}\xspace}                 
 \def\PK      {\ensuremath{K}\xspace}

 \def\Pb      {\ensuremath{b}\xspace}                 
 \def\Pc      {\ensuremath{c}\xspace}

 \def\Pi      {\ensuremath{i}\xspace}

 \def\Pp      {\ensuremath{p}\xspace}

}

\makeatletter
\ifcase \@ptsize \relax
  \newcommand{\miniscule}{\@setfontsize\miniscule{4}{5}}
\or
  \newcommand{\miniscule}{\@setfontsize\miniscule{5}{6}}
\or
  \newcommand{\miniscule}{\@setfontsize\miniscule{5}{6}}
\fi
\makeatother

\DeclareRobustCommand{\optbar}[1]{\shortstack{{\miniscule (\rule[.5ex]{1.25em}{.18mm})}
  \\ [-.7ex] $#1$}}

\def\mup        {{\ensuremath{\Pmu^+}}\xspace}
\def\mun        {{\ensuremath{\Pmu^-}}\xspace} 

\def\cquark    {{\ensuremath{\Pc}}\xspace}

\def\bquark    {{\ensuremath{\Pb}}\xspace}

\def\pion   {{\ensuremath{\Ppi}}\xspace}
\def\piz    {{\ensuremath{\pion^0}}\xspace}

\def\pip    {{\ensuremath{\pion^+}}\xspace}
\def\pim    {{\ensuremath{\pion^-}}\xspace}

\def\kaon    {{\ensuremath{\PK}}\xspace}
  \def\Kbar    {{\kern 0.2em\overline{\kern -0.2em \PK}{}}\xspace}

\def\KorKbar    {\kern 0.18em\optbar{\kern -0.18em K}{}\xspace}

\def\Kp      {{\ensuremath{\kaon^+}}\xspace}
\def\Km      {{\ensuremath{\kaon^-}}\xspace}

  \def\Dbar    {{\kern 0.2em\overline{\kern -0.2em \PD}{}}\xspace}
\def\D       {{\ensuremath{\PD}}\xspace}

\def\DorDbar    {\kern 0.18em\optbar{\kern -0.18em D}{}\xspace}
\def\Dz      {{\ensuremath{\D^0}}\xspace}

\def\Dstarp  {{\ensuremath{\D^{*+}}}\xspace}

\def\Bbar    {{\ensuremath{\kern 0.18em\overline{\kern -0.18em \PB}{}}}\xspace}

\def\BorBbar    {\kern 0.18em\optbar{\kern -0.18em B}{}\xspace}

\def\jpsi     {{\ensuremath{{\PJ\mskip -3mu/\mskip -2mu\Ppsi\mskip 2mu}}}\xspace}

  \def\Y#1S{\ensuremath{\PUpsilon{(#1S)}}\xspace}

\def\proton      {{\ensuremath{\Pp}}\xspace}

\def\Xires       {{\ensuremath{\PXi}}\xspace}

\def\Lbar        {{\ensuremath{\kern 0.1em\overline{\kern -0.1em\PLambda}}}\xspace}
\def\LorLbar    {\kern 0.18em\optbar{\kern -0.18em \PLambda}{}\xspace}

\def\Xibz    {{\ensuremath{\Xires^0_\bquark}}\xspace}
\def\Xibm    {{\ensuremath{\Xires^-_\bquark}}\xspace}

\def\Xicz    {{\ensuremath{\Xires^0_\cquark}}\xspace}

\def\to                 {\ensuremath{\rightarrow}\xspace}

\def\AT#1     {\ensuremath{A_{\mathrm{T}}^{#1}}\xspace}           

\def\C#1      {\ensuremath{\mathcal{C}_{#1}}\xspace}                       
\def\Cp#1     {\ensuremath{\mathcal{C}_{#1}^{'}}\xspace}                    
\def\Ceff#1   {\ensuremath{\mathcal{C}_{#1}^{\mathrm{(eff)}}}\xspace}        
\def\Cpeff#1  {\ensuremath{\mathcal{C}_{#1}^{'\mathrm{(eff)}}}\xspace}       
\def\Ope#1    {\ensuremath{\mathcal{O}_{#1}}\xspace}                       
\def\Opep#1   {\ensuremath{\mathcal{O}_{#1}^{'}}\xspace}                    

\newcommand{\tev}{\ifthenelse{\boolean{inbibliography}}{\ensuremath{~T\kern -0.05em eV}\xspace}{\ensuremath{\mathrm{\,Te\kern -0.1em V}}}\xspace}
\newcommand{\gev}{\ensuremath{\mathrm{\,Ge\kern -0.1em V}}\xspace}
\newcommand{\mev}{\ensuremath{\mathrm{\,Me\kern -0.1em V}}\xspace}
\newcommand{\kev}{\ensuremath{\mathrm{\,ke\kern -0.1em V}}\xspace}
\newcommand{\ev}{\ensuremath{\mathrm{\,e\kern -0.1em V}}\xspace}
\newcommand{\gevc}{\ensuremath{{\mathrm{\,Ge\kern -0.1em V\!/}c}}\xspace}
\newcommand{\mevc}{\ensuremath{{\mathrm{\,Me\kern -0.1em V\!/}c}}\xspace}
\newcommand{\gevcc}{\ensuremath{{\mathrm{\,Ge\kern -0.1em V\!/}c^2}}\xspace}
\newcommand{\gevgevcccc}{\ensuremath{{\mathrm{\,Ge\kern -0.1em V^2\!/}c^4}}\xspace}
\newcommand{\mevcc}{\ensuremath{{\mathrm{\,Me\kern -0.1em V\!/}c^2}}\xspace}

\def\mum  {\ensuremath{{\,\upmu\mathrm{m}}}\xspace}

\def\invfb   {\ensuremath{\mbox{\,fb}^{-1}}\xspace}

\def\ps   {\ensuremath{{\mathrm{ \,ps}}}\xspace}

\def\gsim{{~\raise.15em\hbox{$>$}\kern-.85em
          \lower.35em\hbox{$\sim$}~}\xspace}
\def\lsim{{~\raise.15em\hbox{$<$}\kern-.85em
          \lower.35em\hbox{$\sim$}~}\xspace}

\def\ptot       {\mbox{$p$}\xspace}
\def\pt         {\mbox{$p_{\mathrm{ T}}$}\xspace}

\def\evtgen     {\mbox{\textsc{EvtGen}}\xspace}

\def\geant      {\mbox{\textsc{Geant4}}\xspace}

\def\photos     {\mbox{\textsc{Photos}}\xspace}

\def\pythia     {\mbox{\textsc{Pythia}}\xspace}

\def\tell1  {TELL1\xspace}
\def\ukl1   {UKL1\xspace}

\newcommand{\eg}{\mbox{\itshape e.g.}\xspace}

\usepackage{cite} 
\usepackage{mciteplus}

\usepackage{longtable} 

\def\Xibx    {{\ensuremath{\Xires_\bquark}}\xspace}

\def\XibxPrime  {{\ensuremath{\Xires^{\prime}_\bquark}}\xspace}
\def\XibxStar  {{\ensuremath{\Xires^{*}_\bquark}}\xspace}
\def\XibPrimeMinus  {{\ensuremath{\Xires^{\prime -}_\bquark}}\xspace}
\def\XibPrimeZero  {{\ensuremath{\Xires^{\prime 0}_\bquark}}\xspace}
\def\XibStarMinus  {{\ensuremath{\Xires^{*-}_\bquark}}\xspace}
\def\XibStarZero  {{\ensuremath{\Xires^{*0}_\bquark}}\xspace}

\def\pis  {{\ensuremath{\pi^+}}\xspace}

\newcommand{\kevcc}{\ensuremath{{\mathrm{\,ke\kern -0.1em V\!/}c^2}}\xspace}

\begin{document}

\renewcommand{\thefootnote}{\fnsymbol{footnote}}
\setcounter{footnote}{1}


\begin{titlepage}
\pagenumbering{roman}

\vspace*{-1.5cm}
\centerline{\large EUROPEAN ORGANIZATION FOR NUCLEAR RESEARCH (CERN)}
\vspace*{1.5cm}
\noindent
\begin{tabular*}{\linewidth}{lc@{\extracolsep{\fill}}r@{\extracolsep{0pt}}}
\ifthenelse{\boolean{pdflatex}}
{\vspace*{-2.7cm}\mbox{\!\!\!\includegraphics[width=.14\textwidth]{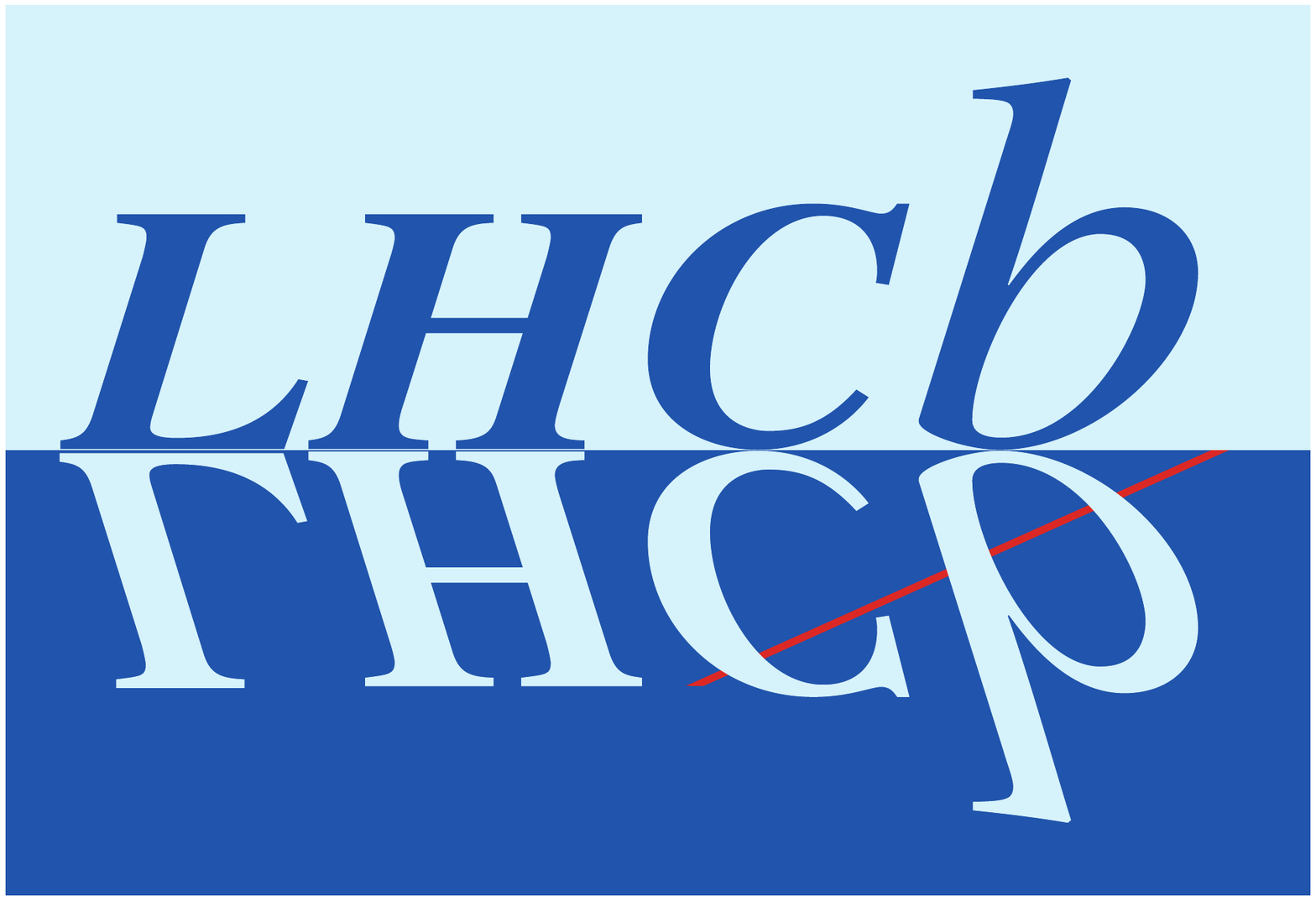}} & &}%
{\vspace*{-1.2cm}\mbox{\!\!\!\includegraphics[width=.12\textwidth]{lhcb-logo.eps}} & &}%
\\
 & & CERN-EP-2016-082 \\  
 & & LHCb-PAPER-2016-010 \\  
 & & 2 June 2016 \\ 
 & & \\
\end{tabular*}

\vspace*{2.0cm}

{\normalfont\bfseries\boldmath\huge
\begin{center}
  Measurement of the properties of the \XibStarZero baryon
\end{center}
}

\vspace*{1.0cm}

\begin{center}
The LHCb collaboration\footnote{Authors are listed at the end of this paper.}
\end{center}

\vspace{\fill}

\begin{abstract}
  \noindent
  We perform a search for near-threshold $\Xibz$ resonances decaying to $\Xibm\pip$
  in a sample of proton-proton collision data corresponding to an integrated luminosity
  of 3\invfb
  collected by the LHCb experiment.
  We observe one resonant state, with the following properties:
  \begin{eqnarray*}
    m(\XibStarZero) - m(\Xibm) - m(\pip) &=& 15.727 \pm 0.068 \, (\mathrm{stat}) \pm 0.023 \, (\mathrm{syst}) \, \mevcc, \\
\Gamma(\XibStarZero) &=& 0.90 \pm 0.16 \, (\mathrm{stat}) \pm 0.08 \, (\mathrm{syst}) \, \mev

    .
  \end{eqnarray*}  
This confirms the previous observation by the CMS collaboration.
The state is consistent with the $J^P=3/2^+$ \XibStarZero resonance
expected in the quark model.
This is the most precise determination of the mass and the
first measurement of the natural width of this state.
We have also measured the ratio
 \begin{align*}
\frac{\sigma(pp\to\XibStarZero X){\cal{B}}(\XibStarZero\to\Xibm\pip)}{\sigma(pp\to\Xibm X)}
 = 0.28 \pm 0.03 \, (\mathrm{stat}) \pm 0.01 \, (\mathrm{syst}) .
\end{align*}
\end{abstract}

\vspace*{2.0cm}

\begin{center}
  Published as JHEP 1605 (2016) 161.
\end{center}

\vspace{\fill}

{\footnotesize 
\centerline{\copyright~CERN on behalf of the \lhcb collaboration, licence \href{http://creativecommons.org/licenses/by/4.0/}{CC-BY-4.0}.}}
\vspace*{2mm}

\end{titlepage}


\newpage
\setcounter{page}{2}
\mbox{~}

\cleardoublepage


\renewcommand{\thefootnote}{\arabic{footnote}}
\setcounter{footnote}{0}



\pagestyle{plain} 
\setcounter{page}{1}
\pagenumbering{arabic}


%

\section{Introduction}

Precise measurements of the properties of hadrons provide important metrics by which
models of quantum chromodynamics 
(QCD), including lattice QCD and potential models employing the symmetries of QCD, can be tested.
Studies of
hadrons containing a heavy quark play a special role since the heavy quark symmetry can be exploited,
for example to relate properties of charm hadrons to beauty hadrons. Measurements of the masses
and mass splittings between the ground and excited states of beauty and charm hadrons provide
a valuable probe of the interquark potential~\cite{Karliner:2008sv}.

There are a number of $b$ baryon states that contain both beauty and strange quarks.
The singly strange states form isodoublets: \Xibz ($bsu$) and \Xibm ($bsd$).
Theoretical estimates of the properties of these states are available
(see, \eg, 
Refs.~\cite{Klempt:2009pi,Karliner:2008sv,Lewis:2008fu,Ebert:2005xj,Liu:2007fg,Jenkins:2007dm,Karliner:2008in,Zhang:2008pm,Wang:2010vn,Brown:2014ena,Valcarce:2008dr,Limphirat:2010zz}).
There are five known \Xibx states which, in the constituent quark model, correspond to five of the
six low-lying states that are neither radially nor orbitally excited:
one isodoublet of weakly-decaying ground states (\Xibz and \Xibm) with $J^P = \frac{1}{2}^+$,
one isodoublet (\XibPrimeZero and \XibPrimeMinus) with $J^P = \frac{1}{2}^+$ but different symmetry properties from the ground states, and
one isodoublet (\XibStarZero and \XibStarMinus) with $J^P = \frac{3}{2}^+$.
The large data samples collected at the Large Hadron Collider have allowed
these states to be studied in detail in recent years.
These studies include precise measurements of the masses and lifetimes of the $\Xibz$ and $\Xibm$ 
baryons~\cite{LHCb-PAPER-2014-021,LHCb-PAPER-2014-048} by the LHCb collaboration, the observation of a peak 
in the $\Xibm\pip$ mass spectrum
interpreted as the 
$\XibStarZero$ baryon~\cite{Chatrchyan:2012ni} by the CMS collaboration, and the observation of two
structures in the $\Xibz\pim$ mass spectrum, consistent with the $\XibPrimeMinus$ and $\XibStarMinus$ 
baryons~\cite{LHCb-PAPER-2014-061} by LHCb.\footnote{
  Charge-conjugate processes are implicitly included throughout.
}
The $\XibPrimeZero$ state was not observed by CMS;
it is assumed to be too light to decay into $\Xibm\pip$.

In this paper, we present the results of a study of the $\Xibm\pis$ mass spectrum,
where the $\Xibm$ baryon is reconstructed through its decay to $\Xicz\pim$, with $\Xicz\to p\Km\Km\pip$.
The measurements use a $pp$ collision data sample recorded by the LHCb experiment, corresponding to
an integrated luminosity of 3\invfb, of which 1\invfb was collected at $\sqrt{s}=7$\tev and 2\invfb at 8\tev.
We observe a single peak in the $\Xibm\pis$ mass spectrum, consistent with the state reported in Ref.~\cite{Chatrchyan:2012ni}.
A precise determination of its mass and the first determination of a non-zero natural width are reported.
We also measure the relative production rate
between the $\XibStarZero$ and $\Xibm$ baryons
in the LHCb acceptance.

The \lhcb detector~\cite{Alves:2008zz,LHCb-DP-2014-002} is a single-arm forward
spectrometer covering the \mbox{pseudorapidity} range $2<\eta <5$,
designed for the study of particles containing \bquark or \cquark
quarks. The detector includes a high-precision tracking system
consisting of a silicon-strip vertex detector surrounding the $pp$
interaction region, a large-area silicon-strip detector located
upstream of a dipole magnet with a bending power of about
$4{\mathrm{\,Tm}}$, and three stations of silicon-strip detectors and straw
drift tubes placed downstream of the magnet.
The tracking system provides a measurement of momentum, \ptot, of charged particles with
a relative uncertainty that varies from 0.5\% at low momentum to 1.0\% at 200\gevc.
The minimum distance of a track to a primary vertex (PV), the impact parameter, is measured with a resolution of $(15+29/\pt)\mum$,
where \pt is the component of the momentum transverse to the beam, in\,\gevc.
Different types of charged hadrons are distinguished using information
from two ring-imaging Cherenkov detectors. 
Photons, electrons and hadrons are identified by a calorimeter system consisting of
scintillating-pad and preshower detectors, an electromagnetic
calorimeter and a hadronic calorimeter. Muons are identified by a
system composed of alternating layers of iron and multiwire
proportional chambers.
The online event selection is performed by a trigger~\cite{LHCb-DP-2012-004}, 
which consists of a hardware stage (L0), based on information from the calorimeter and muon
systems, followed by a software stage, which applies a full event
reconstruction.
The software trigger requires a two-, three- or four-track
secondary vertex which is significantly displaced from
all primary $pp$ vertices and for which the scalar \pt sum of the
charged particles is large. At least one particle should have 
$\pt>1.7\gevc$ and be inconsistent with coming from any of the PVs.
A multivariate algorithm~\cite{BBDT} is used to identify
secondary vertices consistent with the decay of a \bquark hadron.
Only events that fulfil these criteria are retained for this analysis.

In the simulation, $pp$ collisions are generated using
\pythia~\cite{Sjostrand:2006za,*Sjostrand:2007gs} 
 with a specific \lhcb
configuration~\cite{LHCb-PROC-2010-056}.  Decays of hadrons
are described by \evtgen~\cite{Lange:2001uf}, in which final-state
radiation is generated using \photos~\cite{Golonka:2005pn}. The
interaction of the generated particles with the detector, and its response,
are implemented using the \geant
toolkit~\cite{Allison:2006ve, *Agostinelli:2002hh} as described in
Ref.~\cite{LHCb-PROC-2011-006}.

\section{Candidate selection}

\begin{figure}
  \begin{center}
    \includegraphics[width = 0.49\linewidth]{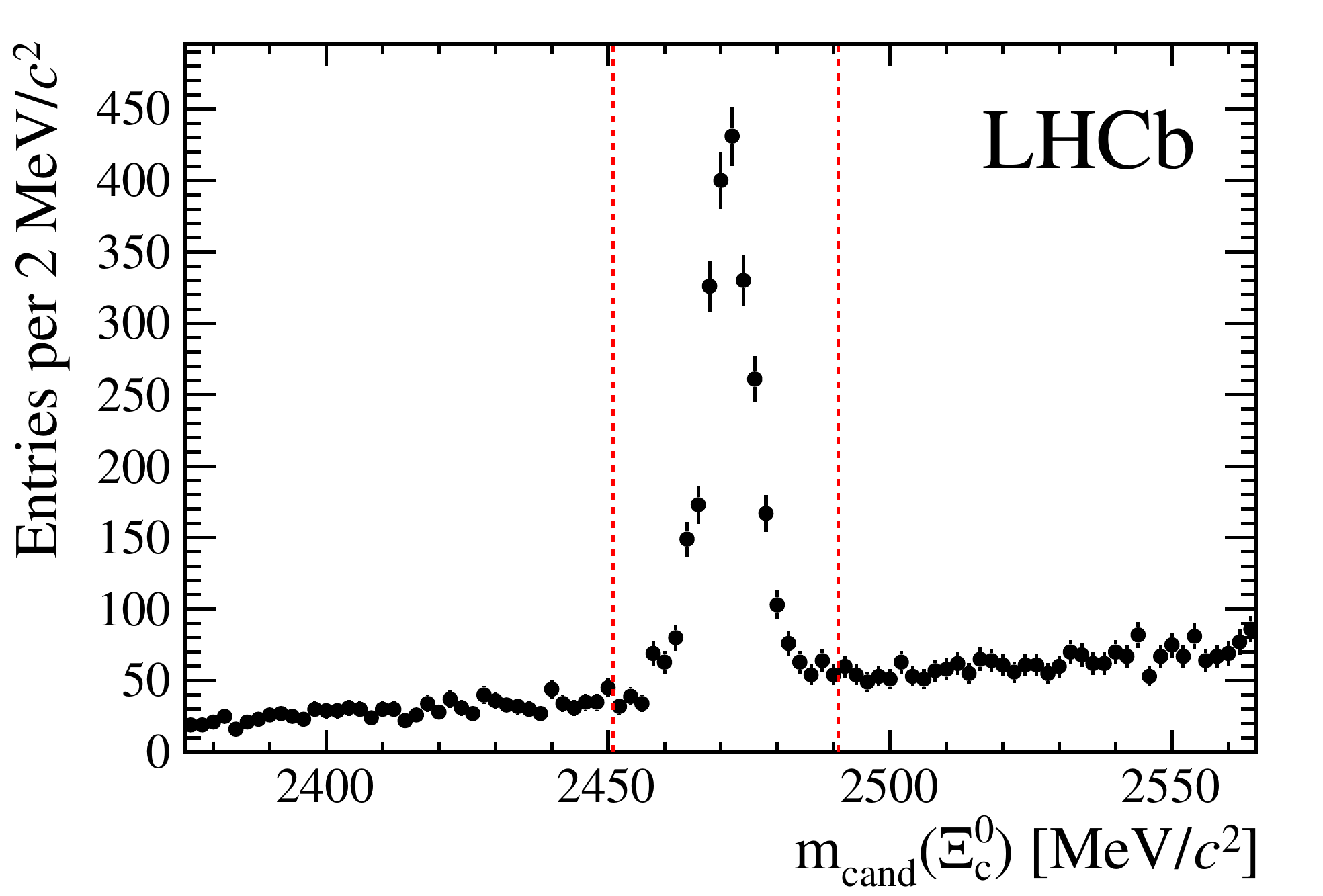}
    \includegraphics[width = 0.49\linewidth]{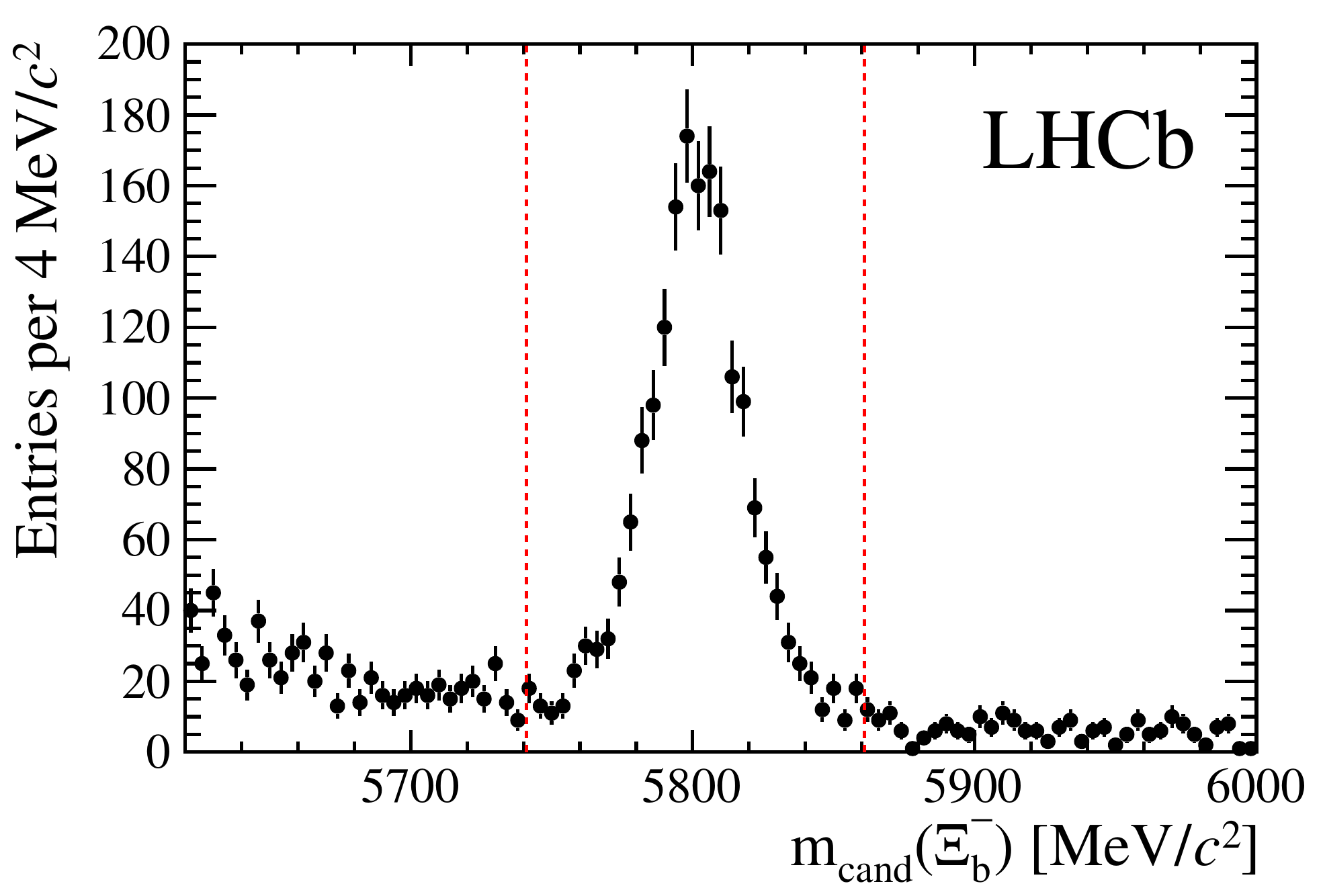}
  \end{center}
  \caption{
    Mass spectra of (left)~\Xicz and (right)~\Xibm candidates
    after all selection requirements are imposed, except for the one on the mass that is plotted.
    The vertical dashed lines show the selection requirements used in forming
    \Xibm and \XibStarZero candidates.
  }
  \label{fig:dataset:massCuts}
\end{figure}

Candidate $\Xibm$ decays are formed by combining $\Xicz\to \proton\Km\Km\pip$ and $\pim$ candidates in a kinematic fit~\cite{Hulsbergen:2005pu}.
All tracks used to reconstruct the $\Xibm$ candidate are required 
to have good track fit quality, have $\pt>100$\mevc, and have particle identification information consistent with the
hypothesis assigned.
The large lifetime of the $\Xibm$ baryon is exploited to reduce combinatorial background
by requiring all of its final-state decay products to have
$\chi^2_{\mathrm{IP}} > 4$ with respect to all of the PVs in the event,
  where $\chi^2_{\mathrm{IP}}$, the impact parameter $\chi^2$,
  is defined as the difference in the vertex fit $\chi^2$ of the
  PV with and without the particle under consideration.
The $\Xicz$ candidates are required to have invariant mass within 
20\mevcc of the known value~\cite{PDG2014}, corresponding to about three times the mass resolution. 
To further suppress background, the $\Xibm$ candidate must
  have a trajectory that points back to one of the PVs ($\chi^2_{\mathrm{IP}} \leq 10$)
  and must have a decay vertex that is significantly displaced from
  the PV with respect to which it has the smallest $\chi^2_{\mathrm{IP}}$
  (decay time $> 0.2$\ps and flight distance $\chi^2 > 100$).
The invariant mass spectra of selected $\Xicz$ and $\Xibm$ candidates are displayed in Fig.~\ref{fig:dataset:massCuts}.

The $\Xibm$ candidates are then required to have invariant mass within 60\mevcc
of the peak value, corresponding to about four times the mass resolution.
In a given event, each combination of \Xibm and \pis candidates is considered, provided that the
pion has \pt greater than 100\mevc and is consistent with coming from the same PV as the $\Xibm$ candidate.
The $\Xibm\pis$ vertex is constrained to coincide with the PV in a kinematic fit,
which is required to be of good quality. The $\Xibm\pis$ system is also required to have 
$\pt>2.5$\gevc.

\begin{figure}
  \begin{center}
    \includegraphics[width = 0.90\linewidth]{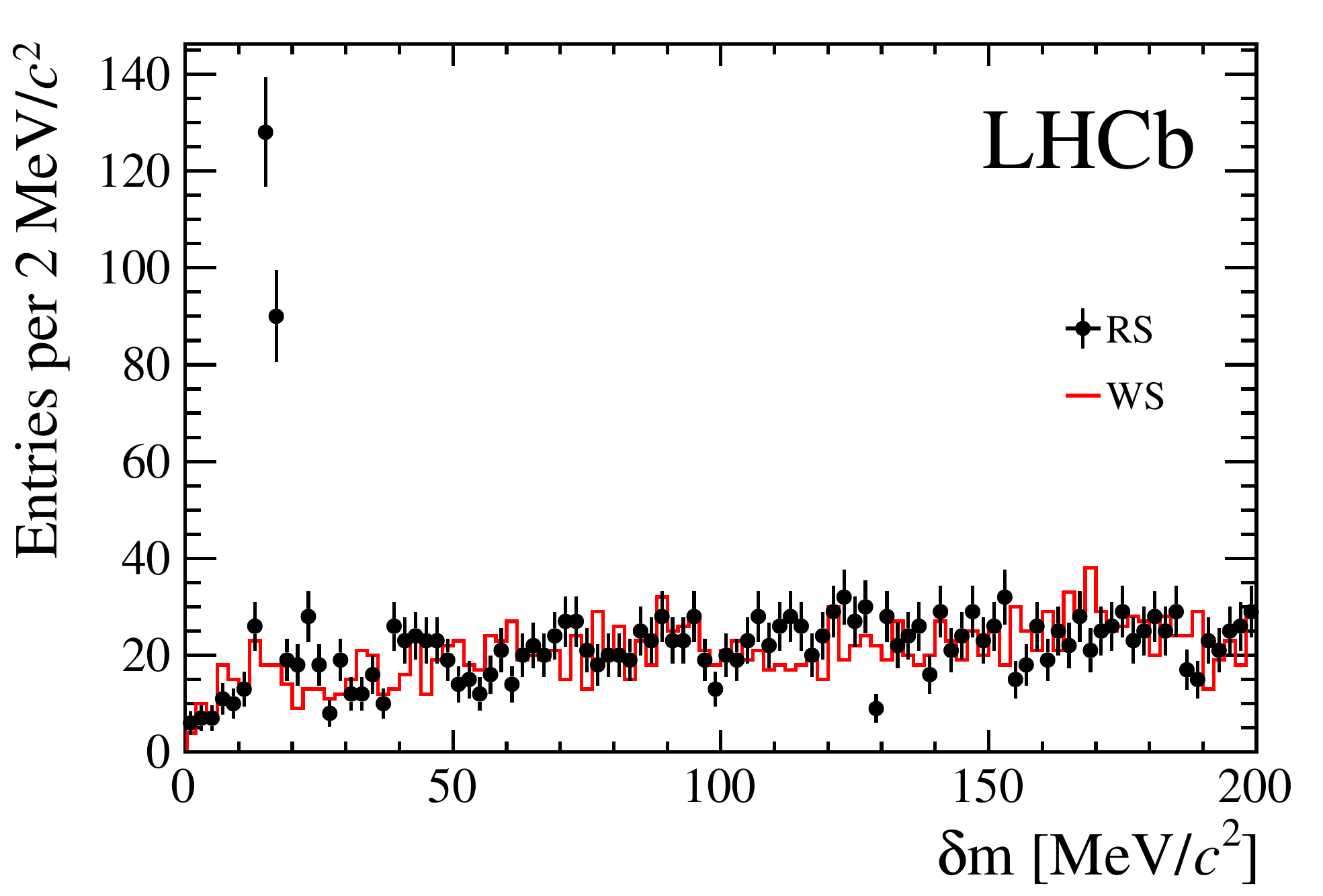}
  \end{center}
  \caption{
    Distribution of $\delta m$.
    Right-sign candidates (RS, $\Xibm \pip$) are shown as points with error bars, and
    wrong-sign candidates (WS, $\Xibm \pim$) as a histogram.
    A single narrow structure is seen in the right-sign data.
  }
  \label{fig:fit:dataRawSpectra}
\end{figure}

The mass difference $\delta m$ is defined as
\begin{align}
  \delta m \equiv m_{\mathrm{cand}}(\Xibm \pip) - m_{\mathrm{cand}}(\Xibm) - m(\pip),
\end{align}
where $m_{\mathrm{cand}}$ represents the reconstructed mass.
The $\delta m$ spectrum of
$\Xibm\pis$ candidates passing all selection requirements is shown in Fig.~\ref{fig:fit:dataRawSpectra}.
A clear peak is seen at about 16\mevcc, whereas no such peak is seen in the wrong-sign ($\Xibm\pim$) combinations, also shown in Fig.~\ref{fig:fit:dataRawSpectra}.

To determine the properties of the $\Xibm\pis$ peak, we consider only candidates with
$\delta m < 45$\mevcc; this provides a large enough region to constrain the combinatorial background shape. 
There are on average 1.16 candidates per selected event in this mass region;
all candidates are kept. 
In the vast majority of events with more than one candidate, a single \Xibm candidate
is combined with different \pis tracks from the same PV.

\section{Mass and width of $\Xibm\pis$ peak}

Accurate determination of the mass, width, and signal yield
requires knowledge of the signal shape, and in particular the mass resolution.
This is obtained from simulated \XibStarZero decays in which the $\delta m$ value is set to
the approximate peak location seen in data.
In this simulation, the natural width of the $\Xibm\pis$ state is fixed to a negligible value so that the shape of the distribution 
measured is due entirely to the mass resolution. The resolution function 
is parameterised as the sum of three Gaussian distributions with a common mean value.
The weighted average of the three Gaussian widths is 0.51\mevcc. 
In the fits to data, all of the resolution shape parameters are fixed to the values obtained from simulation.

Any $\Xibm\pis$ resonance in this mass region would be expected to
have a non-negligible natural width~$\Gamma$. The signal shape in fits to data is therefore described using a
$P$-wave relativistic Breit--Wigner (RBW) line shape~\cite{Jackson:1964zd} 
with a Blatt--Weisskopf barrier factor~\cite{Blatt:1952}, convolved with the resolution
function described above.

The combinatorial background is modelled by an empirical threshold function of the form
\begin{align}
   f(\delta m) = \left( 1 - e^{-\delta m / C} \right) \, (\delta m)^{A},
   \label{eq:bkg}
\end{align}
where $A$ and $C$ are freely varying parameters determined in the fit to the data
and $\delta m$ is in units of \mevcc.

\begin{figure}
  \begin{center}
    \includegraphics[width = 0.98\linewidth]{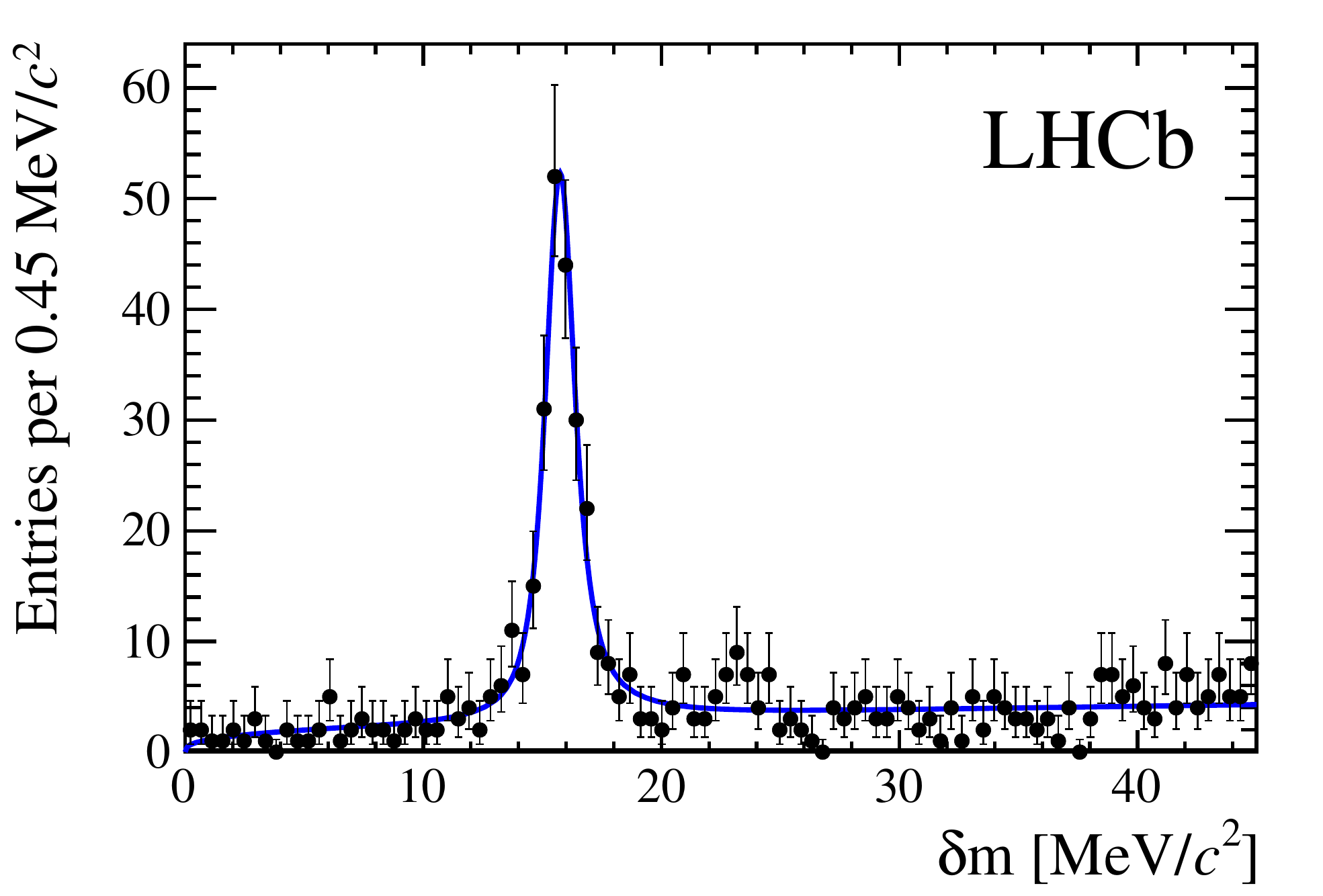}
  \end{center}
  \caption{
    Distribution of $\delta m$ along with the results of the fit described in the text.
  }
  \label{fig:fit:data:fullSpin1BW}
\end{figure}

The mass, width and yield of events in the observed peak are determined from an
unbinned, extended maximum likelihood fit to the $\delta m$ spectrum using the signal and
background shapes described above. The mass spectrum and the results of the fit
are shown in Fig.~\ref{fig:fit:data:fullSpin1BW}.
The fitted signal yield is $232\pm19$ events.
The nonzero value of the natural width of the peak,
$\Gamma = 0.90 \pm 0.16 \mev$ (where the uncertainty is statistical only),
is also highly significant:
the change in log-likelihood when the width is fixed to zero exceeds 30 units.
No other statistically significant structures are seen in the data.

We perform
a number of cross-checks to ensure the robustness of the result.
These include
  splitting the data by magnet polarity,
  requiring that one or more of the decay products of the signal candidate pass the L0 trigger requirements, 
  dividing the data into subsamples in which the \pis candidate has $\pt<250$\mevc and $\pt>250$\mevc,
  varying the fit range in $\delta m$,
  and
  applying a multiple candidate rejection algorithm in which only one candidate, chosen at random, is retained in each event.
In each of these cross-checks, the variation in fit results is consistent with statistical fluctuations.

Several sources of systematic effects are considered and are 
summarised in Table~\ref{tab:sys}.
Other than the first two systematic uncertainties described below, all are determined by making variations to the
baseline selection or fit procedure, repeating the analysis, and taking the
maximum change in $\delta m$ or $\Gamma$.
A small correction (16\kev, estimated with pseudoexperiments) to $\Gamma$ is required due to the
  systematic underestimation of the width in a fit with limited yield;
  an uncertainty of the same size is assigned.
  This correction is already included in the value of $\Gamma$ quoted earlier.
The limited size of the sample of simulated events leads to
  uncertainties on the resolution function parameters. These uncertainties
  are propagated to the final results using the full covariance matrix.
We assign a systematic uncertainty for a particular class of events with
  multiple \XibStarZero candidates in which the \Xibm or \Xicz baryon is misreconstructed.
  This uncertainty is determined by applying a limited multiple candidate rejection
  procedure in which only one \Xibz candidate is accepted per event (but may be
  combined with multiple pions).
The robustness of the resolution model is verified with control samples of
  $\XibPrimeMinus \to \Xibz \pim$ (see Ref.~\cite{LHCb-PAPER-2014-061})
  and $\Dstarp \to \Dz \pip$; based on these tests, the uncertainty is assessed by increasing the
  \XibStarZero resolution width by 11\%. This is the dominant uncertainty on $\Gamma$.
An alternative background description
  is used in the fit to check the dependence
  of the signal parameters on the background model.
The calibration of the momentum scale has an uncertainty of $0.03\%$~\cite{LHCb-PAPER-2012-048,LHCB-PAPER-2013-011},
  the effect of which is propagated to the mass and width of the \XibStarZero baryon.
  As in Ref.~\cite{LHCb-PAPER-2014-061}, this is validated by measuring
  $m(\Dstarp)-m(\Dz)$ in a large sample of $\Dstarp$, $\Dz\to\Km\Kp$
  decays. The mass difference agrees with a recent BaBar
  measurement\cite{Lees:2013uxa,*Lees:2013zna} within 6\kevcc,
  corresponding to $1.3\sigma$ when including the mass scale uncertainty for that decay.
Finally, the dependence of the results on the relativistic Breit--Wigner lineshape 
  is tested: other values of
  the assumed angular momentum (spin 0,\,2) and radial parameter (1--5$\gev^{-1}$) of the
  Blatt--Weisskopf barrier factor are used, and an alternative parameterisation
  of the mass-dependent width (from appendix A of Ref.~\cite{Jackson:1964zd}) is tested.

\begin{table}[bht]
  \caption{
    Systematic uncertainties, in units of \mevcc (mass) and \mev (width).
  }
  \begin{center}
    \begin{tabular}{lcccc}
      Effect & $\delta m$ & $\Gamma$ \\ \hline
      Fit bias correction &  & 0.016 \\
Simulated sample size & 0.007 & 0.034 \\
Multiple candidates & 0.009 & 0.007 \\
Resolution model & 0.001 & 0.072 \\
Background description & 0.002 & 0.001 \\
Momentum scale & 0.009 & 0.001 \\
RBW shape & 0.017 & 0.011 \\
\hline Sum in quadrature & 0.023 & 0.082 \\
\hline Statistical uncertainty & 0.068 & 0.162\\

    \end{tabular}
  \end{center}
  \label{tab:sys}
\end{table}
 
Taking these effects into account, the mass difference and width are measured to be
\begin{eqnarray*}
  m(\XibStarZero) - m(\Xibm) - m(\pip) &=& 15.727 \pm 0.068 \pm 0.023 \mevcc, \\
\Gamma(\XibStarZero) &=& 0.90 \pm 0.16 \pm 0.08 \mev

  ,
\end{eqnarray*}
where the first uncertainties are statistical and the second are systematic.
Given these values, those of the other \Xibx resonances reported previously~\cite{LHCb-PAPER-2014-061},
and the absence of other structures in the $\delta m$ spectrum,
the observed peak is compatible with the $J^P=\frac{3}{2}^+$ state expected in the quark
model~\cite{Klempt:2009pi}, and
we therefore refer to it as the $\XibStarZero$ baryon.

\section{Relative production rate}

In addition to the mass and width of the $\XibStarZero$ state, we measure the rate at which it is
produced in the LHCb acceptance relative to the $\Xibm$ baryon.
The quantity that is measured is
\begin{align}
\frac{\sigma(pp\to\XibStarZero X) \, {\cal{B}}(\XibStarZero\to\Xibm\pis)}{\sigma(pp\to\Xibm X)} = \frac{N(\XibStarZero)}{N(\Xibm)}\frac{1}{\epsilon_{\XibStarZero}^{\rm rel}},
\end{align}
where $\epsilon_{\XibStarZero}^{\rm rel}$ is the ratio of the $\XibStarZero$ to $\Xibm$ selection efficiencies,
and $N$ is a measured yield.
Any variation in the ratio of cross-sections
$\left[ \sigma(pp\to\XibStarZero X) \right] / \left[ \sigma(pp\to\Xibm X) \right]$
between $\sqrt{s}=7$\tev and 8\tev would be far below the sensitivity of our measurements,
and is therefore neglected.

To minimize systematic uncertainties, all aspects of the $\Xibm$ selection are chosen to be
common to the inclusive $\Xibm$ and $\XibStarZero$ samples.
Therefore an additional requirement, not applied to the sample used in the mass and width measurements,
is imposed that at least one of the $\Xibm$ decay products passes the L0 hadron trigger requirements.
The relative efficiency $\epsilon_{\XibStarZero}^{\rm rel}$ includes the efficiency of detecting the
$\pis$ from the \XibStarZero decay and the selection criteria imposed on it.
It is evaluated using simulated decays, and small corrections (discussed below) are applied to account for residual
differences between data and simulation.
Including only the uncertainty due to the finite sizes of the simulated samples, the value of
$\epsilon_{\XibStarZero}^{\rm rel}$ is found to be $0.598\pm0.014$.

\begin{figure}
  \begin{center}
    \includegraphics[width = 0.98\linewidth]{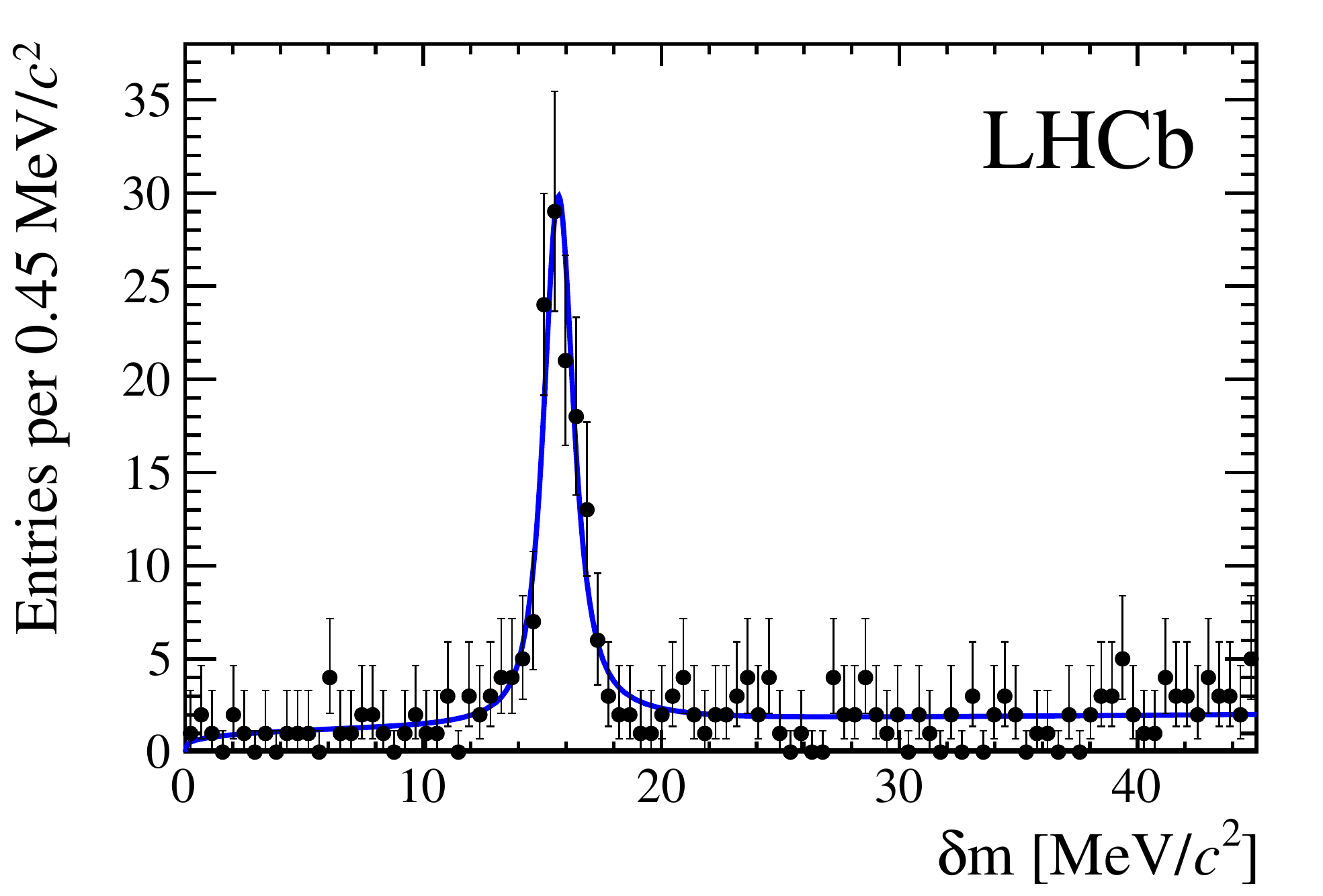}
  \end{center}
  \caption{
    Distribution of $\delta m$, using only events in which one or more of the $\Xibm$ decay products pass the L0 hadron trigger requirements.
    The results of the fit are overlaid.
  }
  \label{fig:fit:data:L0TOS}
\end{figure}

\begin{figure}
  \begin{center}
    \includegraphics[width = 0.98\linewidth]{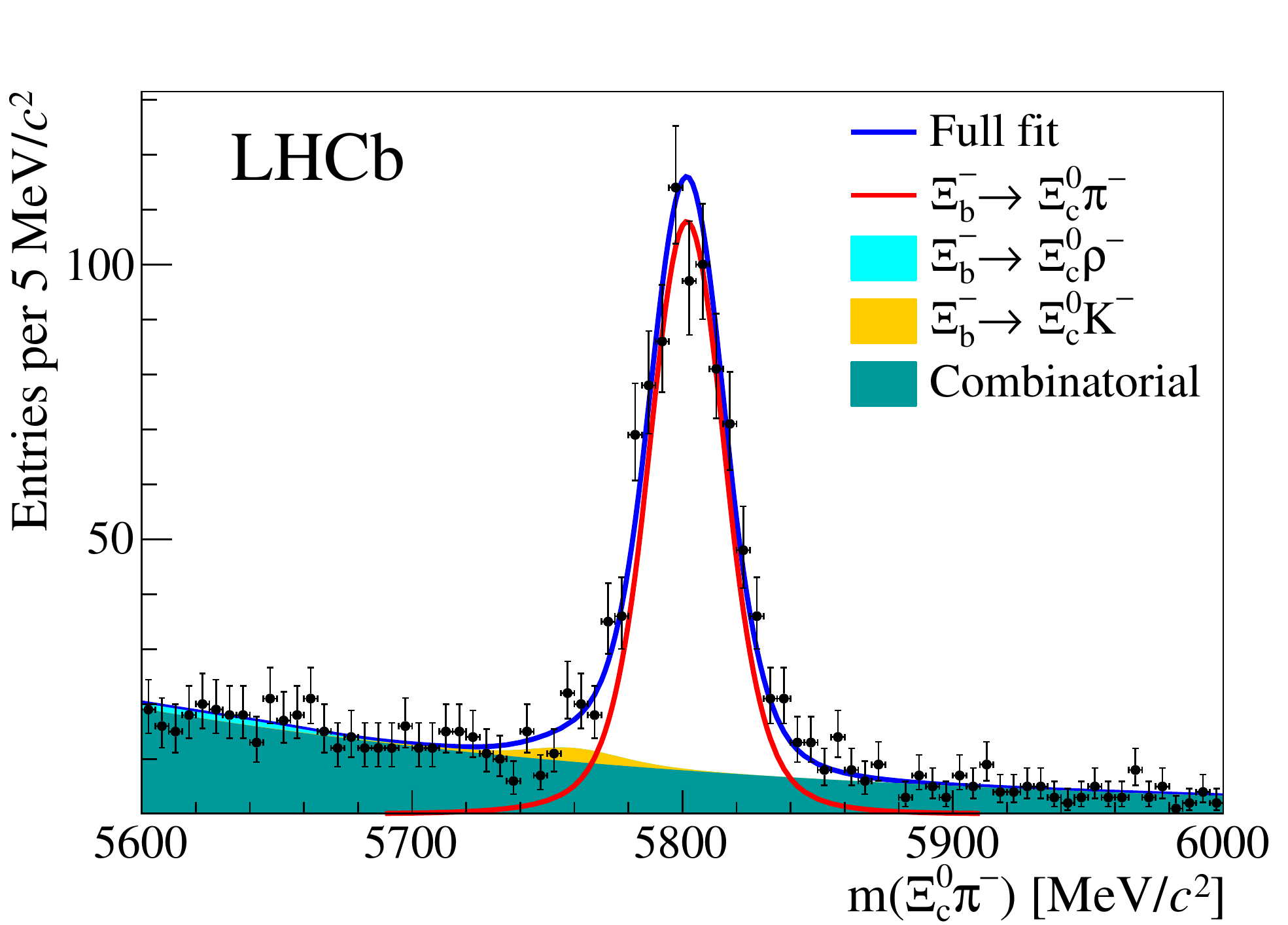}
  \end{center}
  \caption{
    Invariant mass spectrum of selected $\Xicz\pim$ candidates.
    The fit described in the text is overlaid.
    The \Xibm signal peak and background from combinatorial events are clearly
    visible, accompanied by small contributions from the peaking background
    processes $\Xibm \to \Xicz \rho^-$ and $\Xibm\to\Xicz\Km$. 
  }
  \label{fig:Xib2Xic0Pi_withL0}
\end{figure}

The yields in data are obtained by fitting the $\delta m$ and $m_{\mathrm{cand}}(\Xibm)$ spectra
after applying all selection criteria. For the \XibStarZero yield, the data are fitted using the same
functional form as was used for the full sample.
The fit is shown in Fig.~\ref{fig:fit:data:L0TOS}, and the yield obtained is $N(\XibStarZero) = 133\pm14$.
The results of an unbinned, extended maximum likelihood fit
to the \Xibm sample are shown in Fig.~\ref{fig:Xib2Xic0Pi_withL0}.
The shapes used to describe the signal and backgrounds are identical to those 
described in Ref.~\cite{LHCb-PAPER-2014-048}. In brief, the signal shape is described by the sum of two Crystal Ball 
functions~\cite{Skwarnicki:1986xj} with a common mean. The background components are due to misidentified $\Xibm\to\Xicz\Km$ 
decays, partially-reconstructed $\Xibm\to\Xicz\rho^-$ decays, and
combinatorial background. The $\Xibm\to\Xicz\Km$ contribution is also described by the sum of two
Crystal Ball functions with a common mean. Its shape parameters
are fixed to the values from simulation, and the fractional yield relative to that of $\Xibm\to\Xicz\pim$ is fixed to 3.1\%,
based on previous studies of this mode~\cite{LHCb-PAPER-2014-048}.
The $\Xibm\to\Xicz\rho^-$ mass shape is described by an ARGUS function~\cite{Albrecht:1994tb}, 
convolved with a Gaussian resolution function. The threshold and shape parameters are fixed based on simulation,
and the resolution is fixed to 14\mevcc, the approximate mass resolution for signal decays.
The yield is freely varied in the fit. The combinatorial background is
described by an exponential function with freely varying shape parameter and yield.
To match the criteria used for the \XibStarZero selection, only \Xibm candidates within $\pm 60$\mevcc of the known mass 
contribute to the yield, which is found to be $N(\Xibm) = 808\pm32$.

Several sources of uncertainty contribute to the production ratio measurement, either in the signal efficiency or in the
determination of the yields. Most of the selection requirements are common to both the signal and normalization modes, and 
therefore the corresponding efficiencies cancel in 
the production ratio measurement.
Effects related to the detection and selection of
the $\pis$ from the \XibStarZero decay do not cancel, and therefore contribute to the systematic uncertainty.
The tracking efficiency is
measured using a tag and probe procedure with $\jpsi\to\mup\mun$ decays~\cite{LHCb-DP-2013-002}, and for this momentum range
a correction of $(+7.0 \pm 3.0)\%$ is applied.
Fit quality requirements on the \pis track lead to an additional correction of $(-1.5 \pm 1.5)\%$.
The simulation is used to estimate the loss of \XibStarZero efficiency
from decays in which the \pis is reconstructed but has $\pt < 100$\mevc.
This loss, $2.7\%$, is already included in the efficiency, and does not
require an additional correction. Since the simulation reproduces
the \pt spectrum well for $\pt > 100\mevc$, we assign half of the value,
$1.4\%$, as a systematic uncertainty associated with the
extrapolation to $\pt < 100$\mevc.
Finally, the limited sample sizes of simulated events
contribute an uncertainty of $2.4\%$ to the relative efficiency.
With these systematic sources included, the relative efficiency
is found to be $\epsilon_{\XibStarZero}^{\rm rel}=0.598\pm0.026$.

For the \XibStarZero signal yield in data,
we assign a $1\%$ systematic uncertainty due to a potential peaking
background in which a genuine
$\XibStarZero \to \Xibm \pis, \, \Xibm \to \Xicz \pim$ decay is found
but the \Xicz is misreconstructed.
For the normalization mode, independent variations in the signal and background shapes are investigated,
and taken together correspond to a systematic uncertainty in the normalisation mode yield of $2\%$.

Combining the relative efficiency, the yields, and the systematic uncertainties described above, we find
\begin{align} \nonumber
\frac{\sigma(pp\to\XibStarZero X){\cal{B}}(\XibStarZero\to\Xibm\pis)}{\sigma(pp\to\Xibm X)}
 = 0.28 \pm 0.03 \pm 0.01
 ,
\end{align}
where the statistical uncertainty takes into account the correlation between
$N(\XibStarZero)$ and $N(\Xibm)$.

\begin{table}[bht]
  \caption{
    Relative systematic uncertainties on the production ratio.
  }
  \begin{center}
    \begin{tabular}{lccc}
      Effect & Uncertainty \\ \hline
      Simulated sample size & $2.4\%$ \\
      Tracking efficiency correction & $3.0\%$ \\
      Fit quality efficiency correction & $1.5\%$ \\
      Soft pion \pt cut & $1.4\%$ \\
      \XibStarZero yield & $1.0\%$ \\
      \Xibm yield & $2.0\%$ \\ \hline
      Sum in quadrature & $4.9\%$
    \end{tabular}
  \end{center}
  \label{tab:sysProd}
\end{table}

\section{Summary}

Using $pp$ collision data from the LHCb experiment corresponding to an integrated luminosity of 3\invfb,
we observe one highly significant structure in the $\Xibm\pis$ mass spectrum near threshold.
There is no indication of a second state above the $\Xibm\pis$ mass threshold that would indicate
the presence of the $\XibPrimeZero$ resonance; from this we conclude that $m(\XibPrimeZero) \lsim m(\Xibm) + m(\pip)$.
The mass difference and width of the $\XibStarZero$ are measured to be:
\begin{eqnarray*}
  m(\XibStarZero) - m(\Xibm) - m(\pip) &=& 15.727 \pm 0.068 \pm 0.023 \mevcc, \\
\Gamma(\XibStarZero) &=& 0.90 \pm 0.16 \pm 0.08 \mev

  .
\end{eqnarray*}
We interpret the structure as the $J^P=\frac{3}{2}^+$ \XibStarZero state observed previously
by the CMS collaboration through the decay chain $\XibStarZero \to \Xibm \pip, \Xibm \to \jpsi \Xires^-$.
Our results are consistent with and about a factor of ten more precise than their measurements,
$\delta m = 14.84 \pm 0.74 \pm 0.28$\mevcc and
$\Gamma = 2.1 \pm 1.7 \, (\mathrm{stat})\mev$~\cite{Chatrchyan:2012ni}.
The measured width of the state is in line with theory expectations:
a calculation based on lattice QCD predicted a width of $0.51\pm0.16$\mev~\cite{Detmold:2012ge},
and another using the $^3P_0$ model obtained a value of 0.85\mev~\cite{Chen:2007xf}.

Combining our measured value for $\delta m$ with the most precise
measured value of the \Xibm mass,
$5797.72 \pm 0.46 \pm 0.16 \pm 0.26$\mevcc~\cite{LHCb-PAPER-2014-048},
and the pion mass~\cite{PDG2014}, we obtain
 \begin{eqnarray*}
   m(\XibStarZero)  &=& 5953.02 \pm 0.07 \pm 0.02 \pm 0.55 \mevcc

   ,
 \end{eqnarray*}
where the third uncertainty is
due to the $m(\Xibm)$ measurement.
We further combine our result on $\delta m(\XibStarZero)$ with previous LHCb measurements of
$\delta m(\XibStarMinus)\equiv m(\Xibz\pim)-m(\Xibz)-m(\pim)=23.96 \pm 0.12 \pm 0.06 \, \mevcc$~\cite{LHCb-PAPER-2014-061},
and of the ground state isospin splitting, 
$m(\Xibm)-m(\Xibz)=5.92 \pm 0.60 \pm 0.23$\mevcc~\cite{LHCb-PAPER-2014-048}, to obtain the isospin splitting of 
the \XibxStar states,
\begin{align*}
  m(\XibStarMinus) - m(\XibStarZero) &= \delta m(\XibStarMinus) - \delta m(\XibStarZero) - \left[ m(\Xibm)-m(\Xibz) \right] \\
  &= 2.31 \pm 0.62 \pm 0.24 \mevcc 
    .
\end{align*}
In combining the above measurements,
the systematic uncertainties on the mass scale and the RBW shape are
treated as fully correlated between the two $\delta m$ measurements.

We have also measured the inclusive ratio of production cross-sections to be
 \begin{align} \nonumber
\frac{\sigma(pp\to\XibStarZero X){\cal{B}}(\XibStarZero\to\Xibm\pis)}{\sigma(pp\to\Xibm X)}
 = 0.28 \pm 0.03 \pm 0.01 .
\end{align}
This value is similar to the previously measured value from the isospin partner mode, $\XibStarMinus\to\Xibz\pim$,
of
$\displaystyle \frac{\sigma(pp\to\XibStarMinus X){\cal{B}}(\XibStarMinus\to\Xibz\pim)}{\sigma(pp\to\Xibz X)} = 0.21\pm0.03\pm0.01$~\cite{LHCb-PAPER-2014-061}.
Taking into account the neutral modes, e.g. $\XibStarZero\to\Xibz\piz$ and $\XibStarMinus\to\Xibm\piz$, and contributions
from $\XibxPrime$ states~\cite{LHCb-PAPER-2014-061}, it is evident that
in $pp$ collisions at 7 and 8\tev
a large fraction
of $\Xibm$ and $\Xibz$ baryons are produced through feed-down from higher-mass states.









\section*{Acknowledgements}
 
\noindent We express our gratitude to our colleagues in the CERN
accelerator departments for the excellent performance of the LHC. We
thank the technical and administrative staff at the LHCb
institutes. We acknowledge support from CERN and from the national
agencies: CAPES, CNPq, FAPERJ and FINEP (Brazil); NSFC (China);
CNRS/IN2P3 (France); BMBF, DFG and MPG (Germany); INFN (Italy); 
FOM and NWO (The Netherlands); MNiSW and NCN (Poland); MEN/IFA (Romania); 
MinES and FANO (Russia); MinECo (Spain); SNSF and SER (Switzerland); 
NASU (Ukraine); STFC (United Kingdom); NSF (USA).
We acknowledge the computing resources that are provided by CERN, IN2P3 (France), KIT and DESY (Germany), INFN (Italy), SURF (The Netherlands), PIC (Spain), GridPP (United Kingdom), RRCKI and Yandex LLC (Russia), CSCS (Switzerland), IFIN-HH (Romania), CBPF (Brazil), PL-GRID (Poland) and OSC (USA). We are indebted to the communities behind the multiple open 
source software packages on which we depend.
Individual groups or members have received support from AvH Foundation (Germany),
EPLANET, Marie Sk\l{}odowska-Curie Actions and ERC (European Union), 
Conseil G\'{e}n\'{e}ral de Haute-Savoie, Labex ENIGMASS and OCEVU, 
R\'{e}gion Auvergne (France), RFBR and Yandex LLC (Russia), GVA, XuntaGal and GENCAT (Spain), Herchel Smith Fund, The Royal Society, Royal Commission for the Exhibition of 1851 and the Leverhulme Trust (United Kingdom).



\addcontentsline{toc}{section}{References}
\setboolean{inbibliography}{true}
\bibliographystyle{LHCb}
\bibliography{main,LHCb-PAPER,LHCb-CONF,LHCb-DP,LHCb-TDR,REFS}

\newpage

\newpage
\centerline{\large\bf LHCb collaboration}
\begin{flushleft}
\small
R.~Aaij$^{39}$, 
C.~Abell\'{a}n~Beteta$^{41}$, 
B.~Adeva$^{38}$, 
M.~Adinolfi$^{47}$, 
Z.~Ajaltouni$^{5}$, 
S.~Akar$^{6}$, 
J.~Albrecht$^{10}$, 
F.~Alessio$^{39}$, 
M.~Alexander$^{52}$, 
S.~Ali$^{42}$
% ...

{\footnotesize \it
% ...
$ ^{5}$Clermont Universit\'{e}, Universit\'{e} Blaise Pascal, CNRS/IN2P3, LPC, Clermont-Ferrand, France\\
$ ^{6}$CPPM, Aix-Marseille Universit\'{e}, CNRS/IN2P3, Marseille, France\\
% ...
$ ^{10}$Fakult\"{a}t Physik, Technische Universit\"{a}t Dortmund, Dortmund, Germany\\
% ...
$ ^{38}$Universidad de Santiago de Compostela, Santiago de Compostela, Spain\\
$ ^{39}$European Organization for Nuclear Research (CERN), Geneva, Switzerland\\
% ...
$ ^{41}$Physik-Institut, Universit\"{a}t Z\"{u}rich, Z\"{u}rich, Switzerland\\
$ ^{42}$Nikhef National Institute for Subatomic Physics, Amsterdam, The Netherlands\\
% ...
$ ^{47}$H.H. Wills Physics Laboratory, University of Bristol, Bristol, United Kingdom\\
% ...
$ ^{52}$School of Physics and Astronomy, University of Glasgow, Glasgow, United Kingdom\\
% ...
}
\end{flushleft}


\end{document}